\documentclass{PoS}

\title{On the TeV gamma-ray and X-ray correlations exhibited in high-energy peaked BL Lacs: Mrk 501 and 1ES 1959+650.}

\ShortTitle{Correlation between $\gamma$-ray and X-ray emissions in HBLs.}

\author{Mabel Osorio$^{1,2}$;
\speaker{ Rodrigo Sacahui}$^{1}$
; M. Magdalena Gonz{\'a}lez$^{2}$; Nissim Fraija$^{2}$ and Jos{\'e} Andr{\'e}s Garc{\'i}a-Gonz{\'a}lez$^{3}$\\
        $^{1}$ Instituto de Investigaci\'on en Ciencias F\'isicas y Matem\'aticas, ECFM-USAC, Guatemala.\\ 
        $^{2}$Instituto de Astronom\'ia, UNAM, M\'exico\\ 
        $^{3}$Instituto de F\'isica, UNAM, M\'exico\\
        E-mail: \email{jrsacahui@gmail.com, magda@astro.unam.mx}}

\abstract{
 The spectral energy distribution (SED) of  blazars is generally understood through the standard one-zone synchrotron self-Compton (SSC) model, where a strong correlation between X-ray and TeV gamma-ray fluxes is expected. Recently, for Mrk 421 the correlation was confirmed and interpreted within the SSC model arguing that a description of the correlation up to large flux values considerably restricts the value of the magnetic field under the assumption of an electron spectral index independent of the flux state. In this work, we extend the analysis done on Mrk 421 for other two high-energy peaked BL Lacs: Mrk 501 and 1ES 1959+650.
}

\FullConference{36th International Cosmic Ray Conference -ICRC2019-\\
		July 24th - August 1st, 2019\\
		Madison, WI, U.S.A.}

\begin{document}

\section{Introduction}

The broadband spectral energy distribution (SED) of blazars has two well-separated peaks, one of low energy at soft X-rays and the other of high energy at hundreds of GeVs \cite{1997ARA&A..35..445U, 2011ApJ...736..131A}. The SED of blazars is generally understood through the standard one-zone synchrotron self-Compton (SSC) model, where a strong correlation between X-ray and TeV gamma-ray fluxes is expected \cite{2015APh....71....1F,2017ApJS..232....7F}. M.M. Gonz{\'a}lez et al. (2019), tested the correlation between VHE gamma-rays and X-rays fluxes reported by V. A. Acciari et al. (2014) \cite{acciari2014observation} for the blazar Mrk 421. In the analysis, M.M. Gonz{\'a}lez et al. (2019) besides using the Whipple/RXTE data, also included data from HEGRA-CT1/RXTE \cite{aharonian2003tev} and the Milagro Cherenkov telescope. The results were that at larger time scales ($\approx$ months), the correlation was strong, and at smaller time scales ($\approx$ weeks) there was a steeper correlation within 3$\sigma_{d}$, where $\sigma_{d}$ is an intrinsic scatter of the correlation of unknown nature. At high fluxes, independently of the instrument or the time scale of the data, for the larger gamma-ray fluxes the correlation seemed to break down. In the SSC framework, it was found that by having a single and distinctive correlation of gamma-rays and X-rays the values of the magnetic field B can take values in a narrow range of values, making assumptions on parameter space such as flux level dependency on the spectral index. This results served as motivation to extend the study of correlations with other two high-energy peaked BL Lacs (HBLs): Mrk 501 and 1ES 1959+650, both of them presenting a high emission in gamma-rays and X-rays. A preliminary search was done of simultaneous data in gamma-rays and X-rays for both of the sources. In this proceeding it is presented some initial results that will be complemented and interpreted within the SSC framework in the near future.

\section{Markarian 501}
Markarian 501 or Mrk 501 is a nearby BL Lac type blazar at z = 0.034, the second most observed blazar after Mrk 421. It is known to emit bright X-rays and very-high-energy (VHE; E > 100 GeV) gamma-ray photons \cite{quinn1996detection}, for those reasons it has been an object of several studies of multiple broadband, such as Catanese et al. 1997; Kataoka et al. 1999; Petry et al. 2000 and Abdo et al. 2011a (\cite{catanese1997multiwavelength}, \cite{kataoka1999high}, \cite{petry2000multiwavelength}, \cite{abdo2011insights}). 
We present the data of Mrk 501 reported in A. Furniss, et al. (2015) \cite{furniss2015first},
which was taken between April to August of 2013 with the VHE instruments MAGIC and VERITAS for gamma-rays and with the Swift telescope for X-rays. 

\begin{figure}[h]
\centering
\includegraphics[width=0.8\textwidth]{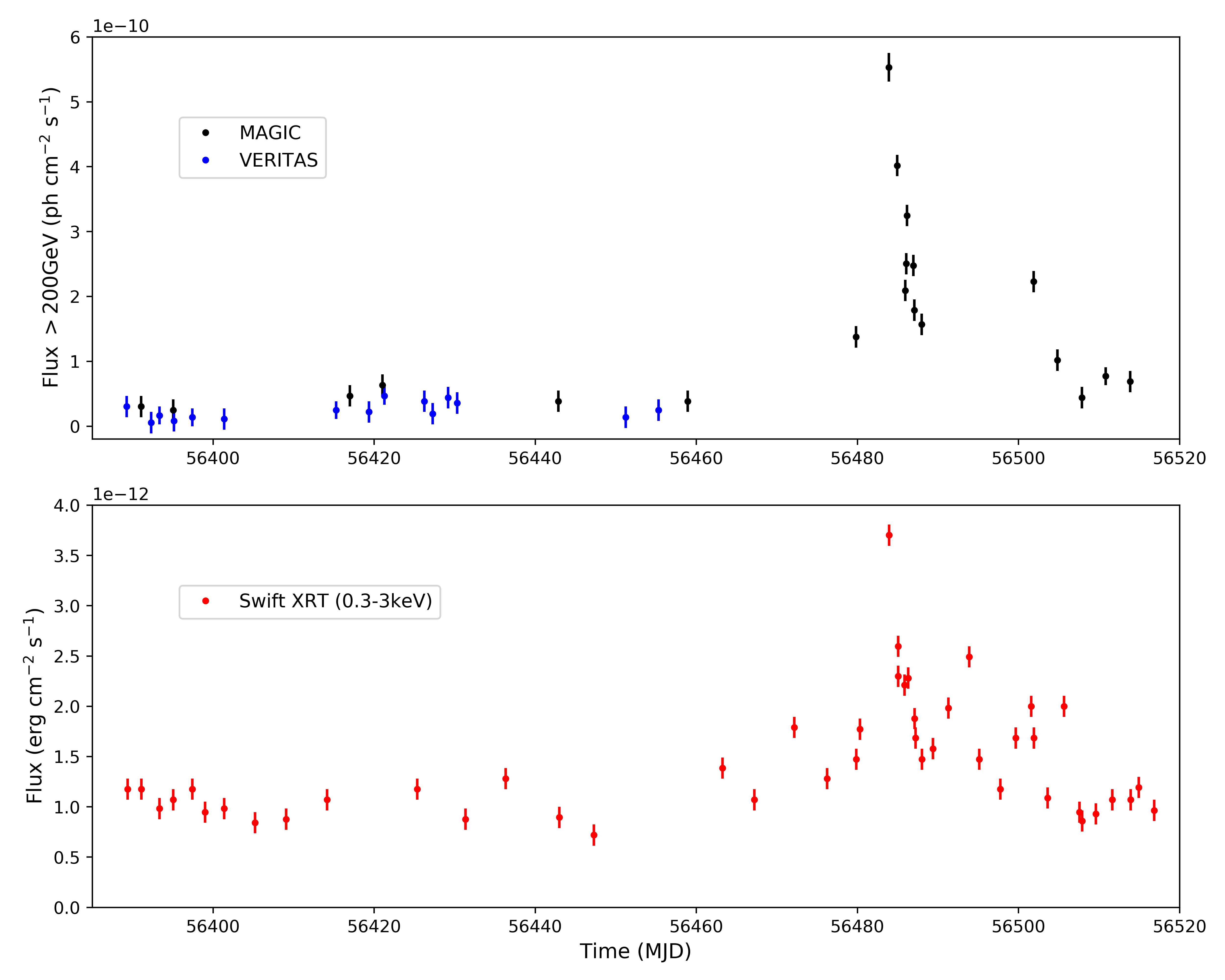}
\caption{Gamma-ray light curve (above) and X-ray light curve (below) of Mrk 501. The data taken was from MAGIC, VERITAS for the gamma-rays and Swift XRT (0.3-3keV) for X-rays. The error bars were estimated by the size of the symbol in the plot reported by A. Furniss, et al.,2015 (see \cite{furniss2015first}), so they are must likely overestimated.}
\label{mkn501}
\end{figure}

MAGIC and VERITAS reported gamma-ray fluxes in units of ph cm$^{-2}$ s$^{-1}$ in energies above 200 GeV. For both instruments a differential power law is fit to the data ($dN/dE=N_0(E/E_0)^{-\Gamma}$) to characterize its VHE spectrum. Their spectral variability shows spectral indexes of $\Gamma$ = 2.50 $\pm$ 0.24 in MAGIC and $\Gamma$ = 3.0 $\pm$ 0.4 in VERITAS in the low state, and present a not so abrupt hardening as it approaches the higher state with spectral index $\Gamma$ = 2.31 $\pm$ 0.05 in MAGIC. VERITAS only observed a quiescent state 
%in a total of fourteen times 
between April 7th (MJD 56389) and June 18th (MJD 56461) of 2013, MAGIC observed the source for a longer time since April 9th (MJD 56391) to August 10th (MJD 56514) during 17 nights (see figure \ref{mkn501}), collecting a total of 22 hours of data and detecting an active state from July 11 (MJD 56484) through July 15th (MJD 56488), reaching fluxes up to 24.3 $\pm$ 0.8 $\times$ 10$^{-11}$ ph cm$^{-2}$ s$^{-1}$ (an order of magnitude higher than in the quiescent state).

For X-rays the Swift XRT satellite was used in the energy interval between 0.3 to 3 keV. It observed Mrk 501 
%59 times 
between January 1st and September 5th (MJD 56293-56540) of 2013. Mrk 501 exhibit a relative constant flux state (see fig \ref{mkn501}) until MJD 56483 when the flux increases to (38.3 $\pm$ 1.5)$\times$ 10$^{-11}$ erg cm$^{-2}$ s$^{-1}$ \cite{furniss2015first}. The data were fit with a power-law model, with index $\Gamma_{pl}$, as well as with a log-parabolic model ($dN/dE=N_0(E/E_0)^{-(\Gamma_{l-p} + \beta log(E/E_0))}$). As the source come closer to its high flux state, the spectral index in both fit models becomes harder from $\bar{\Gamma}_{pl}$ = 2.05 $\pm$ 0.01 in the low state to a nearly constant spectral index with average of $\bar{\Gamma}_{pl}$ = 1.76 $\pm$ 0.03 in the high state. For the log-parabola model, the spectral indexes are from $\bar{\Gamma}_{l-g}$ = 2.06 $\pm$ 0.02 and $\bar{\beta}$ = -0.02 $\pm$ 0.04 for the low state and an average of $\bar{\Gamma}_{l-p}$ = 1.73 $\pm$ 0.06 and $\bar{\beta}$ = 0.08 $\pm$ 0.10 throughout the high state \cite{furniss2015first}.

\section{1ES 1959+650}

1ES 1959+650 is a BL Lac type blazar with a reported redshift of z=0.047 \cite{krawczynski2004multiwavelength}. It is one of the first extragalactic HBL sources detected in the VHE gamma-ray band and in hard X-ray synchrotron emission after Mrk 421 and Mrk 501. We present data of both gamma-ray and X-ray emissions reported in \cite{krawczynski2004multiwavelength}. For gamma-ray data, 1ES 1959+650 was monitored by the Whipple Cherenkov and for X-rays the RXTE satellite was used.

\begin{figure}[h]
\centering
\includegraphics[width=0.8\textwidth]{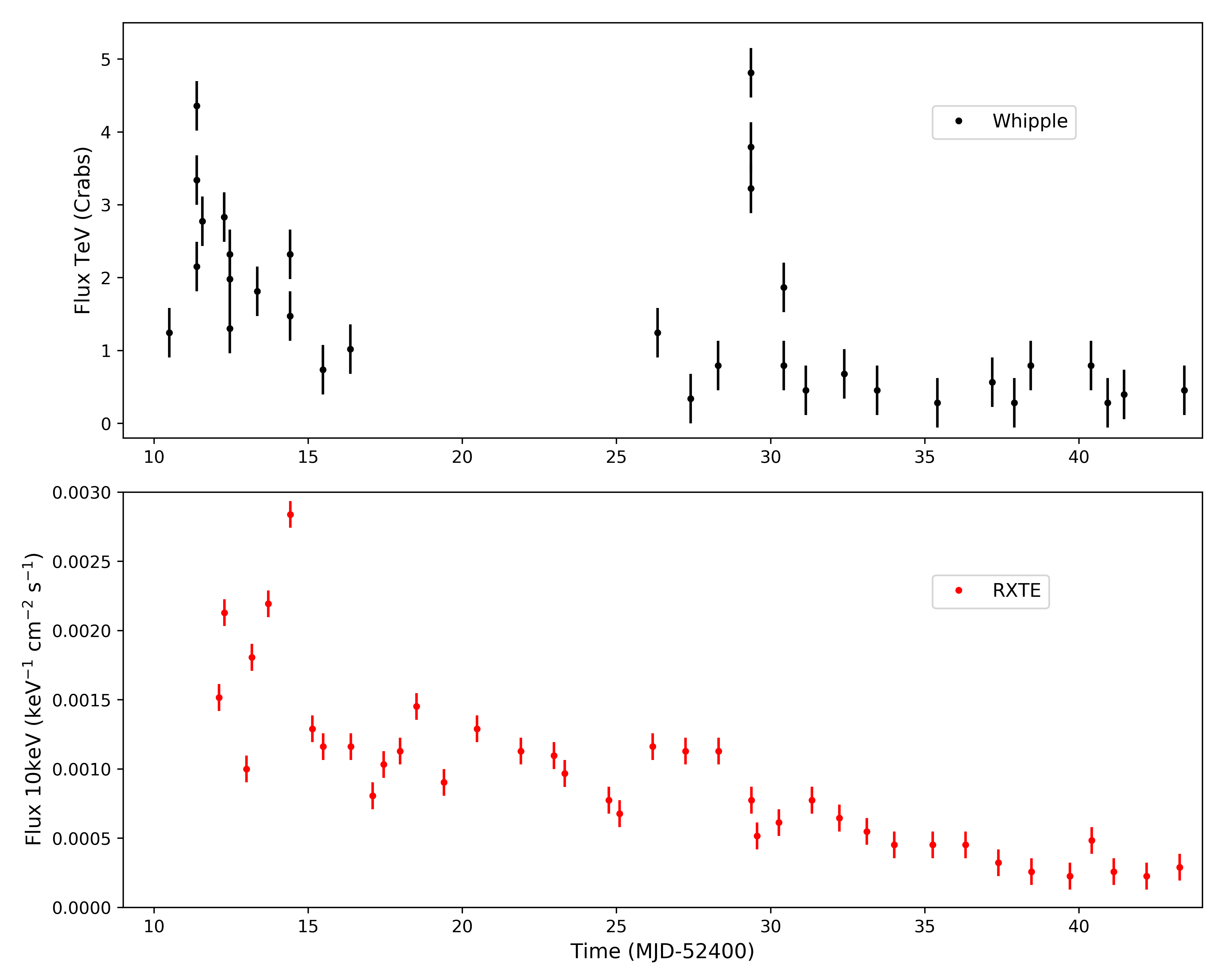}
\caption{The plot above shows the gamma-ray light curve observed by Whipple at 600 GeV and below the X-ray light curve observed by RXTE in 10 keV of HBL 1ES 1959+650. The error bars were estimated by the size of the symbol in the plot reported by H. Krawczynski, et al., 2004 \cite{krawczynski2004multiwavelength}, so they are must likely overestimated.}
\label{1ES}
\end{figure}

The Whipple telescope observed the source in energies beyond 600 GeV and measured the flux in Crab units. The observations were made from May 16th to July 8th of 2002. The total data collected with Whipple consisted of 39.3 hours of on-source data. Because of the location of the Whipple telescope and the source, some corrections had to be made for large zenith angles involving a temporary reduction of the telescope detection efficiency. For this reason the energy spectra could not be determined with the standard tools (see \cite{krawczynski2004multiwavelength}). Detection of strong flares were observed between May 16th and 17th of 2002 and an orphan gamma-ray flare was reported by Whipple on June 4th 2002 (see figure \ref{1ES}).

The X-ray data was obtained from the Proportional Counter Array (PCA) on board the RXTE satellite \cite{Jah06}, it uses units of keV$^{-1}$ cm$^{-2}$ s$^{-1}$ and measures at energies of 10 keV. The time of observation was between May 16th and August 14th of 2002 with a total exposure in the time intervals from 160 s to 4.43 ks. For the spectral model a single-power-law function was a suitable fit for all the data.

Between May 16th and 17th of 2002 (MJD 52410-52411) a strong flare was observed in gamma-rays with Whipple as well as in X-rays by RXTE, in this time interval, both emissions shows correlation but around the 4th of June (MJD 52429) the orphan gamma-ray flare was observed without a counterpart in X-rays (see fig \ref{1ES} \cite{krawczynski2004multiwavelength}), for this reason, the correlation was calculated only in the low state.

\section{Analysis}

The maximum likelihood approach is used to estimate the robustness of the correlations and to know how significant are data deviations from it. The data is assumed to be affected by an intrinsic scatter $\sigma_{d}$ of unknown nature quantified using the maximum likelihood method as discussed by D'Agostini, 2005 \cite{d2005fits}. This method states that if the correlation (F$_{\gamma, i}$, F$_{X,i}$) behaves linearly, i.e. $F_{\gamma} = aF_{X}+b$, with an intrinsic scatter $\sigma_{d}$, the best values of the parameters (a, b and $\sigma_{d}$) can be determined by minimizing the minus-log-likelihood function, in which the uncertainties of F$_{\gamma, i}$ and F$_{X,i}$ (the standard deviation of each) are taken into account. 

\ 

This method was used in the work made by M.M. Gonz{\'a}lez et al. 2019 \cite{gonzalez2019reconcilement}, the correlation remained until a break-down at high gamma-ray fluxes at all the different time scales studied. In the month-time scale, the data in the region of low VHE gamma-rays of the correlation presented a smaller dispersion (within 3$\sigma_{d}$) of the best fit but still there was couple of points of higher fluxes out of the correlation. In the week and hour-time scale the correlation had out layers further than 3$\sigma_{d}$, with respect to the month-time scale fit, especially for the highest gamma-ray fluxes. This could be explained as the existence of other physical processes producing the highest gamma-ray fluxes.  With the two new sources studied here we will test if this behavior only belongs to Mrk 421 or if is a proper behavior of other HBLs. 

% that is, fluxes over longer time scales (i.e. months) were dominated predominantly by the quiescent states, while fluxes over shorter time scales (i.e. weeks to hours) were measured in flaring states because of the instrument sensitivities. 

\section{Results and Conclusions}

\begin{figure}[h]
\centering
\includegraphics[width=0.8\textwidth]{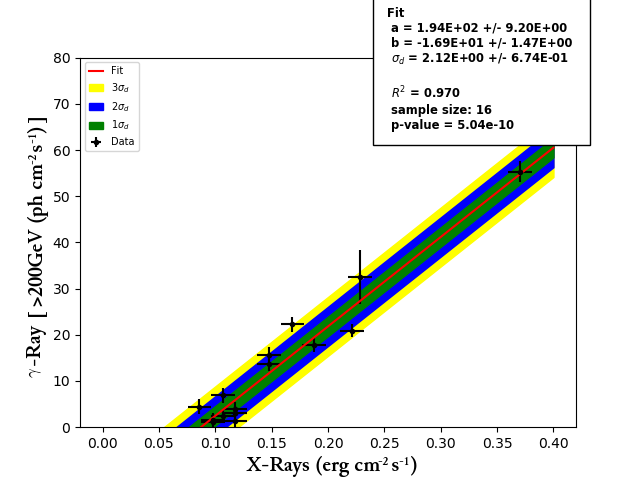}
\caption{Correlation of gamma-rays and X-rays for Mrk 501 with data of MAGIC and VERITAS for the gamma-rays and the Swift telescope for the X-rays. The data lack time simultaneity in an average of 30 minutes. With these data there is no evidence of a break-down at high fluxes, all of the points are within 3$\sigma_{d}$.}
\label{mkn501_corr}
\end{figure}

\begin{figure}[h]
\centering
\includegraphics[width=0.8\textwidth]{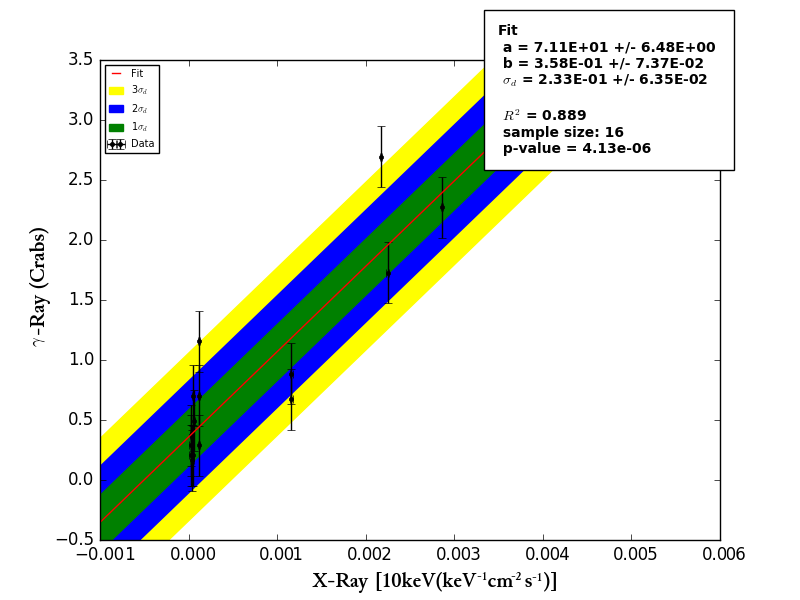}
\caption{Correlation of gamma-rays and X-rays for the HBL 1ES 1959+650 with data of Whipple for gamma-rays and of RXTE for X-rays for only a low state. The data is not simultaneous in an average time of 1.7 hours, however the correlation seems not to be affected by it.}
\label{1ES_corr}
\end{figure}

Using the data obtained from MAGIC-VERITAS/Swift for Mrk 501 by \cite{furniss2015first} and from Whipple/RXTE for 1ES 1959+650 by \cite{krawczynski2004multiwavelength}, we got the correlation of gamma-rays and X-rays. The plots are shown in figures \ref{mkn501_corr} and \ref{1ES_corr} respectively. The parameters of the fit for Mrk 501 are a = 194$\pm$9.20, b = -16.9$\pm$1.47, $\sigma_{d}$ = 2.12$\pm$0.67 with a p-value of 5.04$\times$10$^{-10}$ and a Pearson correlation factor of R$^2$ = 0.970. For 1ES 1959+650 the parameters are a = 71.10$\pm$6.48, b = 0.36$\pm$0.07, $\sigma_{d}$ = 0.23$\pm$0.06 with a p-value = 4.13$\times$10$^{-6}$ and R$^2$ = 0.889. With these preliminary results, we notice the lack of a break-down at high gamma-ray fluxes and the description of the data with a linear correlation from low to high gamma-ray fluxes. 

%poner unidades en las graficas, quitar la parte de "datos" fig 3 4. el error bar es el tamaño del puntero, las incertidumbres son probablemente sobreestimados. algunos sí tienen errores. con los resultados si hay evidencia que haya correlacion pero no hay evidencia de divergencia del break down a flujos altos. puede ese comportamiento sea solo de mkn 421, se necesitan mas estudios.
%poner en el caption la simultaneidad 
%sacar el promedio de simultaneidad de los tiempos
%poner caption
%para mkn 501 y 1ES qué tan simultaneos son los datos (ver en las tablas de datos digitalizados y sacar un promedio) (preguntar a rodrigo por los datos de 1ES)

%\section{Conclusions}
\ 

The article presented by M.M. Gonz{\'a}lez et al., 2019 \cite{gonzalez2019reconcilement} served as a motivation to make an analysis of correlation between gamma-rays and X-rays of the HBLs Mrk 501 and 1ES 1959+650 which are sources of high emission in both energy ranges. The data used was reported by \cite{furniss2015first} and \cite{krawczynski2004multiwavelength}, respectively, and the correlation analysis was made with a method of maximum likelihood described by G. D'Agostini, 2005 \cite{d2005fits}. \\

In the current initial results, we notice that the correlations exist even though there is not an exact time simultaneity of the gamma-ray and X-ray observations, or whether the spectrum is variable as in  both sources. Likewise, the correlations do not show out layers at high gamma-ray fluxes, it might be that the behavior of Mrk 421 can not be generalized to other HBLs or that our data set is reduced. In the near future a bigger data sample will be obtained for both sources, the effect of the quasi-simultaneity will be studied and the theoretical model to find a range of values for their magnetic field will be applied.

\section*{Acknowledgements}
The authors would like to thank DGAPA-UNAM for financial support by grants AG100317, IA102019 and DIGED-USAC.

%aumentar conjunto de datos, estudiar el efecto de simultanedad. No parece que la pseudosimultaneadad afecte la correlacion. 
%el espectro constante de 1Es es constante, el de Mkn 501 es variable pero en ambos casos la correlacion no es afectada 

%ver de 1ES 
%Arreglar la lista de autores y el titulo

\end{document}